\documentclass[epj]{svjour}
\usepackage{graphics}
\usepackage{hyperref}
\usepackage{epsf}
\usepackage{float}
\usepackage{amsmath,amssymb,amscd}
\usepackage{mathrsfs}
\usepackage{color}
\usepackage{subfigure}
\usepackage{cancel}

\newcommand{\be}{\begin{eqnarray}}
\newcommand{\ee}{\end{eqnarray}}
\newcommand{\beq}{\begin{equation}}
\newcommand{\eeq}{\end{equation}}

\newcommand{\nn}{\nonumber}

\newcommand{\pio}{\mbox{$\pi^0$}}
\newcommand{\gzero}{g_0^\theta}
\newcommand{\gone}{g_1^\theta}
\newcommand{\mstar}{m_\ast}
\newcommand{\btheta}{\bar\theta}
\newcommand{\mn}{m_N}
\newcommand{\mpi}{M_{\pi}}
\newcommand{\fpi}{F_{\pi}}
\newcommand{\Mstr}{\delta m_{np}^{\rm str}}
\newcommand{\Mpistr}{(\delta\mpi^2)^{\rm str}}
\newcommand{\Mnp}{\delta m_{np}}
\newcommand{\NthreeLO}{{\sf N$^3$LO}}
\newcommand{\NtwoLO}{{\sf N$^2$LO}}

\newcommand{\NLO}{{\sf NLO}}
\newcommand{\LO}{{\sf LO}}
\newcommand{\CPv}{\cancel{\rm CP}}
\newcommand{\Pv}{\not{\rm P}}
\newcommand{\Tv}{\not{\rm T}}

\newcommand{\unitmatrix}{\mathbf{1}_2}
\newcommand{\half}{{\textstyle\frac{1}{2}}}
\newcommand{\third}{{\textstyle\frac{1}{3}}}

\definecolor{grey}{rgb}{0.3,0.3,0.3}

\usepackage{amsmath}
\usepackage{amssymb}
\usepackage{epsfig}

\newcommand{\mev}{\,\mbox{MeV}}

\begin{document}

\title{The electric dipole moment of the deuteron from the QCD {$\theta$}-term}

\author{J.~Bsaisou\inst{1}
      \and C.~Hanhart\inst{1}\fnmsep\inst{2}\fnmsep\inst{3}
      \and  S.~Liebig\inst{1}
      \and U.-G.~Mei{\ss}ner\inst{1}\fnmsep\inst{2}\fnmsep\inst{3}\fnmsep\inst{4}\fnmsep\inst{5}
      \and A.~Nogga\inst{1}\fnmsep\inst{2}\fnmsep\inst{3} 
      \and A.~Wirzba\inst{1}\fnmsep\inst{2}\fnmsep\inst{3}}

\institute{Institut f\"{u}r Kernphysik  and J\"ulich Center for Hadron Physics, 
           Forschungszentrum J\"{u}lich,  D-52425 J\"{u}lich, Germany
                 \and Institute for Advanced Simulation,  Forschungszentrum J\"{u}lich,  D-52425 J\"{u}lich, Germany
            \and JARA --  Forces And Matter Experiments, Forschungszentrum J\"{u}lich,  D-52425 J\"{u}lich, Germany
            \and  Helmholtz-Institut f\"ur Strahlen- und Kernphysik, Universit\"{a}t Bonn, D-53115 Bonn, Germany
            \and  Bethe Center for Theoretical Physics,  Universit\"{a}t Bonn, D-53115 Bonn, Germany
             }

\date{\today}

\abstract{
The  two-nucleon contributions to the electric dipole moment (EDM) of
the deuteron, induced by the QCD $\theta$-term,  are calculated in the
framework of effective field theory up-to-and-including next-to-next-to-leading
order. In particular we find
for the difference of the deuteron EDM and the sum of proton and
neutron EDM induced by the QCD $\theta$-term a value of 
$(-5.4\pm 3.9)\,\bar\theta\times 10^{-4}\, e$\,fm. The by far dominant uncertainty
comes from the CP- and isospin-violating $\pi NN$ coupling constant.
} 

\maketitle

\section{Introduction} \label{sec: intro}
Under the assumption that the CPT theorem is valid,
permanent electric dipole moments (EDMs) of elementary particles and nuclei, which arise
under parity P {\em and}  time-reflection T breaking,  belong to the most 
promising signals of CP-violating physics beyond the Cabibbo--Kobayashi--Maskawa (CKM) phase of
the Standard Model (SM)\,\cite{Khriplovich_Lamoreaux,Bigi_Sanda,Pospelov_Ritz}. Possible mechanisms\,\cite{Baluni,Crewther}
are the dimension-four  $\theta$ vacuum angle term of
Quantum Chromodynamics (QCD)\,\cite {'tHooft:1976up}
and the effective dimension-six  quark, quark-color,  and gluon-color terms 
\cite{Buchmuller:1985jz,DeRujula:1990db,Grzadkowski:2010es} (including
certain combinations of four-quark terms \cite{Maekawa:2011vs,deVries_dim_six})  resulting from extensions of the SM  such as 
supersymmetry\,\cite{RamseyMusolf_Su},
many-Higgs scenarios\,\cite{Weinberg:1989dx} etc.
In refs.~\cite{mannel1,mannel2} it was recently pointed out that the same mechanism
that drives the potential CP violation beyond the SM in $D\to
K^+K^-/\pi^+\pi^-$~\cite{LHCb,CDF} should, if present, also lead to an enhanced nucleon EDM.
 However,
a single successful measurement of an EDM signal of the neutron, say, 
would not suffice to isolate the specific CP-violating
mechanism. Therefore, more than one EDM measurement involving other hadrons and (light) nuclei, {\it e.g.}
the proton, deuteron, helium-3, are necessary in order to uncover the
source(s) of the CP breaking.

In recent years various theoretical studies focussed on the calculation of
EDMs for light
nuclei~\cite{Khriplovich:1999qr,Lebedev_Olive,Liu_Timmermans,stetcu,deVries:2011re,deVries:2011an,Afnan:2010xd,Gibson:2012zz},
largely triggered by on-going plans for dedicated experiments to measure
EDMs of light ions using storage rings~\cite{Semertzidis:2003iq,Orlov:2006su,BNL_deuteron,Lehrach,JEDI}.
These calculations revealed that different \ CP-violating\ me\-cha\-nisms 
contribute to different probes with different strength.
Therefore, non-zero measurements as well as controlled calculations of nucleon and
nuclear EDMs are necessary to reveal additional information on the physics
beyond the SM that drive non-vanishing EDMs. 

In this work we calculate the two-nucleon contribution to the deuteron EDM
 that would be produced from
a non-vanishing QCD $\theta$-term up to next-to-next-to-leading  order. Thus,
once the EDMs of the proton, neutron and deuteron were measured, the results
of our calculation would allow one to extract the value of $\bar \theta$
directly from data, assuming that no other CP-violating mechanisms contribute significantly.
Since lattice QCD will eventually be able to calculate the neutron and proton
EDMs with  $\bar \theta$ as the only input, a combination of the calculation presented here
with lattice QCD  and experimental numbers will enable  one to decide, if the
$\theta$-term is  the culprit of generating the EDMs. 
Note that direct lattice 
calculations for nuclear EDMs would be
much more challenging.
\begin{figure*}  
  \begin{center}
  \includegraphics[width=14.0cm]{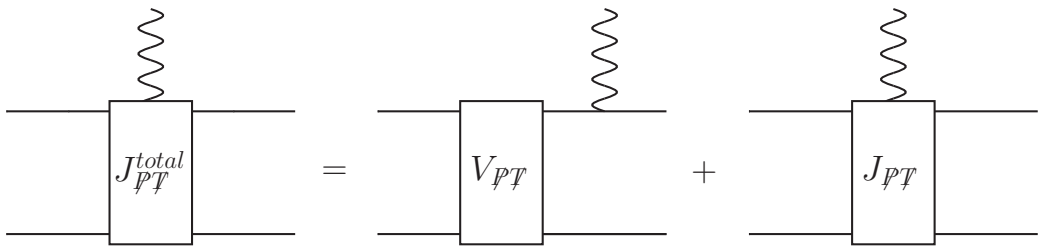}
   \caption[Total CP-violating transition current]
   {Total CP-violating transition current. The P- and T-violation stem  from either CP-violating 
   two-nucleon potentials or two-nucleon-irreducible CP-violating  transition currents. \label{fig: total_current}}
  \end{center}
\end{figure*}

The terms of the CP-violating interaction Lagrangian relevant for this work are given
by --- see ref.~\cite{deVries:2011an} and references therein ---
\begin{eqnarray}
   \mathcal{L}_{\CPv}&= &\bar N\left(b_0 + b_1 \tau_3\right) S^\mu
  N v^\nu F_{\mu\nu} 
     + g_0\bar{N}\vec{\pi}\cdot\vec{\tau}N+g_1\bar{N}\pi_3N \nonumber \\
&&+ C_1^0\bar{N}ND_{\mu}(\bar{N}S^{\mu}N)
  +C_2^0\bar{N}\vec{\tau}ND_{\mu}(\bar{N}\vec{\tau}S^{\mu}N)\nonumber \\
&&+C_{1}^3\bar{N}\tau_3ND_{\mu}(\bar{N}S^{\mu}N)
+C_{2}^3\bar{N}ND_{\mu}(\bar{N}\tau_3S^{\mu}N)\nonumber \\
&&+\cdots\,.
\label{eq: lth}
\end{eqnarray}
Here $v^\mu = (1,\vec 0)$, $S^\mu =(0,\half\vec\sigma)$  and
$\vec\tau$
are the nucleon velocity, spin and isospin, respectively, while $D_\mu$ is the covariant derivative. 
For $\theta$-term-induced CP violation naive dimensional analysis (NDA) gives that
$\gone/\gzero\sim \epsilon M_\pi^2/m_N^2$ \cite{Mereghetti:2010tp}. However,
as already pointed out in ref.~\cite{Lebedev_Olive} and refined further below,
$\gone$ is significantly enhanced compared to this estimate --- in fact, the
contribution from $\gone$ dominates the deuteron EDM.

The single-nucleon EDM from the $\theta$-term starts to
contribute at the one loop level \cite{Baluni,Crewther}.
At the same order there are two counter
terms, proportional to $b_0$ and $b_1$, to absorb the
divergence \cite{Pich,Ottnad,Ottnad_diplom}  --- for a recent update see ref.\,\cite{GuoMei2012}.  Therefore, although the value of the CP-violating coupling
constants $g_0$ and $g_1$ can be related to the strength of the QCD $\theta$-term, $\bar \theta$, 
within the effective field theory the same is not possible
for the EDM of  a {\em single} neutron or proton. This is different in case of the nuclear EDMs:
for the {\em few-nucleon} contributions,  counter terms appear  only at subleading orders and
therefore controlled calculations become feasible, although,
in case of the deuteron EDM, with a sizable uncertainty.
Such subleading terms can   be found in the second and third lines of eq.~(\ref{eq: lth}), where the two terms in third line are additionally suppressed by isospin breaking. 
 The dots in  eq.~(\ref{eq: lth}) denote further CP-violating terms that do not contribute to the deuteron EDM at orders considered in this work.
 These terms include 
CP-violating $NN\pi\gamma$-,  $NN \pi\pi$-,
 \mbox{$NN\pi\pi\gamma$-,}\ $4N\gamma$-terms,  CP-violating photon--two-pion terms, and CP-violating pure pion terms (see ref.~\cite{Mereghetti:2010tp} for the latter class).

There are two types of contributions that are relevant for the present study,
namely CP-violating $NN$ interactions and CP-violating irreducible $NN\to
NN\gamma$ transition currents --- {\it c.f.} fig.~\ref{fig: total_current}.
As will be outlined below, for the deuteron EDM the latter kind of
contributions contains at its leading non--vanishing order loop diagrams that
are calculated in this work for the first time.

The paper is structured as follows: in sect.~\ref{sec:piNNCPv} the prefactors of the 
CP-violating $\pi NN$ couplings $\gzero$ and $\gone$ are derived from the QCD $\theta$-term.
After this, a brief discussion of the
power counting is presented in sect.~\ref{power}.
Section\,\ref{sec: thetaEDM} contains the derivation of the two-nucleon contributions to the deuteron  EDM
induced by the $\theta$-term, where the $NN$ potential and transition current contributions are discussed
in subsections \ref{sec: NN_potential} and \ref{sec: Transition_Current}, respectively. 
Finally, in sect.~\ref{sec: Conclusions} a short summary of the presented results and an outlook are
given.  The role that the vacuum alignment plays for the generation of $\gone$ is outlined in appendix~\ref{app: vacuum}.
Appendices~\ref{app: lebedev} and \ref{app: su3} present  two further alternatives  to  derive  the CP-violating coupling constant
$\gone$, an update of the original derivation 
by Lebedev \textit{et al.}\,\cite{Lebedev_Olive} and a derivation in the framework of SU(3)
chiral perturbation theory (ChPT), similarly
to the  one of $\gzero$ by \cite{Crewther,Pich,Ottnad,Ottnad_diplom}, respectively. Finally, appendix \ref{app: odd parity} is reserved for an estimate
of the $\gone$ contribution resulting from a resonance-saturation mechanism involving the odd-parity nucleon-resonance $S_{11}(1535)$.

\section{CP-violating {\boldmath$\pi NN$} couplings from the {\boldmath$\theta$}-term} \label{sec:piNNCPv}

On the quark level the effect of the $\theta$-term can be written
as $ \mstar\bar \theta \bar q i\gamma_5 q$~\cite{Crewther}, with the reduced quark mass
$\mstar \equiv m_u m_d/(m_u{+}m_d)$ = $(m_u{+}m_d)(1{-}\epsilon^2)/4$, where 
$\epsilon$ = $(m_u-m_d)/(m_u+m_d)$ = $-0.35\pm 0.10$~\cite{pdg}.
 It thus behaves under chiral 
rotations identically to the quark mass term and can be
included in the chiral Lagrangian via
\begin{equation}
    \chi_\pm = u^\dagger \chi u^\dagger \pm u\chi^\dagger u  
\quad \mbox{with}
    \ \chi=2B(s+ip) \ ,
   \label{eq: chi_pm}
\end{equation}
where $s$ may for our purposes be identified with the quark mass matrix,
which reads
\begin{equation}
{\cal M} = \frac{m_u+m_d}{2}\,\unitmatrix + \frac{m_u-m_d}{2}\tau_3
\label{massmatrixSU2}
\end{equation}
 while $p=\mstar\bar\theta\, \unitmatrix$. 
 The pion fields are contained in the usual SU(2) matrix  $u = U^{1/2}$, see {\it e.g.}  \cite{BKM1995}. 

Starting point for the calculation of the CP-violating $\pi NN$ vertices
are, to the order we are working,
the quark mass dependent terms of the CP-conserving Lagrangian  ${\cal
  L}^{(2)}_{\pi N}$~\cite{BKM1995}, namely
\begin{eqnarray} \nonumber
  \lefteqn{c_1N^\dagger\langle \chi_+\rangle N +
  c_5N^\dagger\left[\chi_+-\frac12\langle \chi_+\rangle\right] N} \\ \nonumber
  && \quad = \phantom{\  \,  \mbox{}+} c_1 \ 4B(m_u{+}m_d) \ N^\dagger N \\
  &&\quad\ \  \,  \mbox{}+  \ c_5 \ 2B \ N^\dagger\left[(m_u{-}m_d)\tau_3+\frac{2\mstar\bar\theta}{F_\pi}(\vec \pi\cdot \vec \tau)\right]N 
  \nonumber\\
 && \quad\ \  \,  \mbox{} + \ \cdots\,.
  \label{terms}
\end{eqnarray}
Here $\langle \cdot \rangle$ denotes the trace in flavor space.
The dots indicate that terms not relevant for this study were omitted.

We start with a discussion of the term proportional to $c_5$.
The first term of the third line  of eq.~(\ref{terms}) leads to the quark-mass-induced part
of the proton--neutron mass difference, $ \Mstr$. It can be quantified
from three different sources: (i) the use of dispersion theory
to quantify the electromagnetic part of the proton--neutron mass
difference~\cite{Cini,Cottingham,Gasser:1982ap,Walker-Loud}, (ii) lattice QCD~\cite{beane}, or (iii)
from charge-symmetry-breaking (CSB) studies of $pn\to d\pi^0$~\cite{CSBpn2dpi}. 
All analyses lead to consistent results, with the first one being the most
accurate. Thus we will use~\cite{Walker-Loud} 
\begin{equation}
   4B(m_u-m_d)c_5 = \Mstr = (2.6\pm 0.5 )\, \mbox{MeV} \ .   
\end{equation}
{}From this we get
\begin{equation}
    \gzero=  \frac{\Mstr (1 -\epsilon^2)}{4 F_\pi \epsilon}\, 
     \btheta\, =( -0.018 \pm 0.007)\,\bar\theta \ ,  
     \label{gzero_form}
\end{equation}
where we used $F_\pi  =  92.2\,{\rm MeV}$ \cite{pdg}.
The superscript $\theta$ indicates that we here  only include the strength that
comes from the $\theta$-term.  The expression given above agrees with the
prediction of ref.~\cite{Mereghetti:2010kp} when eq.~(14) of
ref.~\cite{Mereghetti:2010kp} is inserted into the corresponding
eq.~(8)\footnote{Note that the result of ref.~\cite{Mereghetti:2010kp} has the opposite sign to 
   ours (which is compensated by the opposite
  sign of $\epsilon$). Furthermore, $F_\pi $ is defined  twice as large there.}. 
It turns out that the value of $\gzero$ is more than a factor of 10 smaller
than the estimate from NDA given by $\bar\theta
M_\pi^2/(m_N F_\pi)$ in terms of the pion mass $M_\pi$, the nucleon mass $m_N$ and the pion axial
decay constant $F_\pi$.

The first term in the second line of eq.~(\ref{terms}) leads to the 
quark-mass-induced isoscalar contribution to the nucleon mass --- thus $c_1$ can be related
to the $\pi N$ sigma term. For this low-energy constant (LEC) we use the value given in ref.~\cite{pid},
\beq c_1=(-1.0\pm
   0.3)\,{\rm GeV}^{-1} \ ,
\eeq
which is a compilation of various extractions of $c_1$
\cite{BL01,GR02,BM99,FM00}. At this stage this contribution does not
contain a CP-odd term, however, as outlined in ref.~\cite{Mereghetti:2010tp}
and detailed within our formalism in appendix \ref{app: vacuum}, in the
presence of CP violation a rotation of the vacuum is necessary in order
to remove pion tadpoles from the theory. This rotation induces an additional
CP-violating term in the pion--nucleon sector; namely, in agreement with
 ref.~\cite{Mereghetti:2010tp} we find a coupling of $g_1$ type:
\begin{equation}
  \gone=\frac{2\,c_1\,\Mpistr\,(1-\epsilon^2)}{\fpi\,\epsilon}\btheta\, ,
  \label{eq: gone first}
 \end{equation}
where $\Mpistr$ denotes the
 quark-mass-induced part
of the mass-square-splitting between  charged and neutral pions.
Note that the above-mentioned vacuum rotation  produces as well a correction to $\gzero$,
which, however, is numerically negligible.

\begin{figure}[t!]  
	\centering
   \includegraphics[scale=0.8]{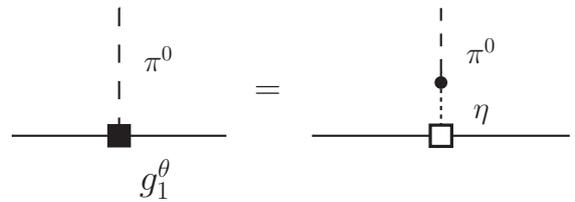}
  \caption[CP-violating pi NN vertex]
 { 
  CP-violating $\pi^0 NN$ vertex  $g_1^{\theta}$  (black square) induced by $\pio$--$\eta$-mixing and the 
  CP-violating $\eta NN$   vertex (open square).} 
   \label{fig: pioeta}
\end{figure}

 Inserting the relation
 \cite{GassLeut_eta}
\begin{equation}
     \Mpistr \approx\frac{B}{4}\frac{(m_u{-}m_d)^2}{m_s{-}(m_u{+}m_d)/2}\approx\frac{\epsilon^2}{4}\frac{\mpi^4}{M_K^2-\mpi^2}
\end{equation}
  into eq. (\ref{eq: gone first})
we get the result
 \begin{equation}
\gone \approx \frac{c_1(1-\epsilon^2)\epsilon}{2\fpi} \,\frac{\mpi^4}{M_K^2-\mpi^2}\,\btheta
 = (0.003\pm 0.001) \,\bar \theta \, ,
  \label{eq: gone second}
\end{equation}
where the uncertainty of this contribution is dominated by the uncertainty in $c_1$.  The expression given in (\ref{eq: gone second}) exactly agrees with the one 
presented in appendix \ref{app: lebedev} which is derived from $\eta$--$\pio$ mixing, see ref.\,\cite{Lebedev_Olive} and fig.\,\ref{fig: pioeta},
 provided the strange-quark content of the nucleon is vanishingly small. An alternative derivation, which uses SU(3) ChPT input instead of
sigma-term estimates, is presented
in appendix~\ref{app: su3}. Taking the rather large SU(3) errors into consideration,
 the SU(3) estimates for $\gone$  (and $\gzero$) are  compatible with our final values  which are quoted at the end of this section.

In addition to the contribution related to the $\pi N$ sigma term there exists
one additional, linearly independent operator structure that leads
to a contribution to $\gone$, see ref.\,\cite{Mereghetti:2010tp}. In our notation, it is
given by
\begin{equation} 
\frac{c_1^{(3)}}{4}N^\dagger\langle \chi_-\rangle^2 N =
c_1^{(3)}\frac{B^2m^*(m_u{-}m_d)}{\fpi}\bar{\theta}N^\dagger\pi_3 N + \cdots \ . 
\label{chimi2}
\end{equation} 
Unfortunately, this operator structure contributes
to CP-conserving observables at too high order that it could be constrained
from a study of, say, $\pi N$ scattering. Thus   we need to 
estimate the value of $c_1^{(3)}$ differently.
While the  operator $\chi_+$ leads to terms that are even (odd) in the pion field
for CP-conserving (violating) contributions, these relations are inverted for the operators $\chi_-$:  
CP-conserving (violating) contributions are given by terms that are odd (even) in the pion field.
 Thus, a natural
resonance saturation estimate for the operator of eq.~(\ref{chimi2})
 is given by a diagram, where one insertion of $\chi_-$ converts
the even-parity nucleon into the lowest odd-parity  nucleon-resonance,
the $S_{11}(1535)$, which then decays via an isospin-violating decay
into a neutral pion and a nucleon. The latter step may be modeled by
a   $S_{11}(1535)$ decaying into $\eta N$ which then converts into
$\pi^0 N$ via $\eta-\pi$ mixing. This contribution is potentially
important, since the coupling of this nucleon resonance to $\eta N$
is very significant~\cite{pdg}. However, an explicit calculation, see appendix \ref{app: odd parity}, shows
that  the mentioned contribution does not exceed the value estimated from
NDA. 
Moreover, in order to get the proper SU(3) chiral limit of QCD,
the $\eta$ should be coupled with a derivative even to the nucleon resonances ---
the resulting Lagrangian is given in ref.~\cite{achot} --- which leads
to an additional suppression. We therefore consider it save to estimate
the additional   $\gone$-uncertainty due to our ignorance of $c_1^{(3)}$  from an NDA estimate which is equal 
to $\epsilon \mpi^4/(m_N^3 \fpi)\sim 0.002$. In what follows
we will therefore use
\begin{equation}
\gone = (0.003\pm 0.002) \bar \theta \, ,
\end{equation}
which includes zero within two sigma.
In particular, we find for the ratio
\begin{equation}
  \frac{\gone}{\gzero} = \frac{8c_1\Mpistr }{ \Mstr }
    =  - 0.2 \pm 0.1\,. 
    \label{our_ratio}
\end{equation}
The value of $\gone/\gzero$
 is numerically about a factor of 25 larger than the SU(2)
estimate of order $\epsilon M_\pi^2/m_N^2$, 
which would follow from the first relation of eq.\,(\ref{our_ratio}) if the
scaling   $\Mstr \sim \epsilon \mpi^2/m_N$ were assumed. The main origin of this difference is  that 
$g_0^\theta$ is unusually small --- instead of two powers
in the counting the relative suppression numerically is of the 
order of one power in the expansion parameter $M_\pi/m_N$. It is this observation that
we will use in the power counting as outlined in the next section.

\section{Power counting}
\label{power}

\begin{figure*}[t!]  
	\centering
 {\includegraphics[width=0.75\textwidth]{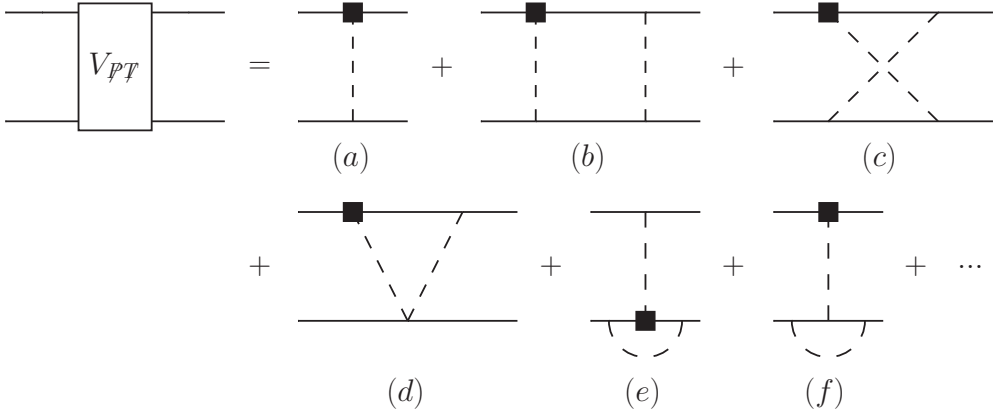}}
  \caption[CP-violating two-nucleon potentials]
  {
   Contributions to the CP-violating two-nucleon potential: (a)
   \LO\ contributions, (b)-(f) \NLO\  and \NtwoLO\ contributions, where the
   former class contains the $\gzero$ and the latter the $\gone$ coupling. 
 Solid lines denote nucleons and dashed
lines denote pions. The CP-violating vertex is depicted by a black box. For each class of diagrams only one representative is shown.
    \label{fig:potential}}
\end{figure*}

It is crucial for this study to identify a power counting that allows a comparison of 
the contributions to the nuclear EDMs from CP-odd transition
currents to those from the CP-odd $NN$ potential. The  power counting
originally proposed by Weinberg for nuclear matrix
elements~\cite{weinbergpiNN},  in spite of its many successful applications, is not able to explain analogous ratios studied
numerically in ref.~\cite{liebig} --- we will therefore modify it
slightly, as explained below. An alternative scheme is presented in
ref.~\cite{deVries:2011an}.

In Weinberg's counting, contributions to the deuteron EDM that come from a CP
violating
potential ({\it c.f.} fig.~\ref{fig: total_current}) are regarded as reducible,
while the transition currents are counted as irreducible. Thus, one needs
to power-count the nuclear wave functions and the photon couplings separately,
making it necessary to assign a scale to a disconnected nucleon line. 
For dimensional reasons the corresponding $\delta^{(3)}$ function is identified with $1/p^3$, where $p$
denotes the typical momentum
appearing
in the evaluation of the integrals, identified with the pion mass, $M_\pi$.
However, if indeed nucleon momenta are of order $M_\pi$, the two-nucleon
intermediate state appearing between the photon coupling and the CP-violating
$NN$ potential is off-shell. Thus, also this contribution is to be
regarded
as irreducible with the two-nucleon propagator counted as $(p^2/m_N)^{-1}$,
where $m_N$ denotes the nucleon mass. Again $p$ is identified with $M_\pi$.
This power counting properly explains the numerical observations of
ref.~\cite{liebig} and will be used in this work as well.
For more details we refer to ref.~\cite{piAinprep}.

\subsection{Power counting for the contributions of the single-nucleon EDMs}

In a world where CP violation beyond the SM is driven by the $\theta$-term,
within the effective field theory the single-nucleon EDMs start at the one-loop
level. At the same order there are two counter terms --- the $b_i$ terms
in eq.~(\ref{eq: lth}). The isospin structure of the loops gives that 
the isoscalar component of the single-nucleon EDMs is suppressed by one 
order in the counting compared to the isovector one~\cite{Baluni,Crewther}.
However, this suppression is not present for the counter
terms~\cite{Pich,Ottnad}, and therefore for the power counting we may estimate
both the contribution from the $d_0$ as well as the $d_1$ term from the
estimate for the leading
loop contribution given by
 \begin{eqnarray*}
&& \gzero  \times (M_\pi/F_\pi)\times (e M_\pi) \times (1/M_\pi^5)\times (M_\pi)^4/(4\pi)^2\\
 &&\sim e\gzero
F_\pi M_\pi/m_N^2 \ ,
\end{eqnarray*}
where the dimension-full factors in the first line come from the regular $\pi NN$ vertex, the
photon-pion vertex (with  the electron charge $e<0$), the propagators and the integration measure,
respectively, and we identified $(4\pi F_\pi)\sim m_N$. In order to derive
from this the total transition current we need to multiply the estimate
with $(1/F_\pi^2) \times m_N/M_\pi^2$ from the $NN$ potential and the two-nucleon
propagator, respectively. We therefore find  an estimate of the order of
$e\gzero/(F_\pi m_N M\pi)$ from the single-nucleon EDM
for the leading contribution
to the total transition current. Thus, the single-nucleon EDMs start to
contribute to the deuteron EDM at \NLO, as we will outline in the next
subsections --- {\it c.f.} table~\ref{tab: PowerCounting}.
\begin{table*}[thb]
   \caption[Power-counting scales]
   {Power-counting scales of the  CP-violating $NN$ potentials (left) and 
    (total) transition currents (right)  relevant for the two-nucleon contribution
   to the  $\theta$-term-induced EDM of the deuteron. 
   Note that the equivalence  $4\pi F_\pi \sim m_N$ is assumed.\label{tab: PowerCounting}
                   }
                   \centering
   \begin{tabular}
  {l  l c l    l c l}  
 \hline\noalign{\smallskip}
     &   \multicolumn{3}{c}{$NN$ potential\ \ \ \ \ \ } & \multicolumn{3}{c}{(total) transition current}\\ 
     \noalign{\smallskip}\hline\noalign{\smallskip}
    \LO & $\gzero/(\mn F_\pi)$ &$\sim$& $\gone /(\mpi F_\pi) $   & $\gzero\,e/(\mpi^2 F_\pi)$& $\sim$ & $\gone\,e\,\mn /(\mpi^3 F_\pi)$ \\
    \NLO &$\gzero \mpi/(\mn^2 F_\pi)$ &$\sim$ &$\gone/(\mn F_\pi)$ &  $\gzero\,e/(\mpi \mn F_\pi)$& $\sim$ &$\gone\,e/(\mpi^2 F_\pi)$  \\
    \NtwoLO\ \ \ \ \  &$\gzero \mpi^2/(\mn^3 F_\pi)$& $\sim$& $\gone \mpi /(\mn^2 F_\pi)$ \ \ \ \ \ \ \ &  $\gzero\,e/(\mn^2 F_\pi)$& $\sim$ &$\gone\,e/(\mpi\mn F_\pi)$ \\
  \noalign{\smallskip}\hline
\end{tabular}
\end{table*}

\subsection{Power counting of the irreducible CP-odd {\boldmath $NN$} potential}

The leading diagrams  for the irreducible CP-odd $NN$ potential
are shown in fig.~\ref{fig:potential}.
The leading, isospin-conserving, CP-odd one-pion exchange can be  
estimated as $\gzero /(M_\pi F_\pi)$. However, as will be discussed in the next section, 
this term does not contribute to the deuteron EDM due to selection
rules. The first non-vanishing contribution comes from the subleading,
isospin- and CP-odd coupling $\gone$.
It is
estimated to contribute as $\gone/(M_\pi F_\pi)\sim \gzero/(m_N F_\pi)$, where
we used the empirical relation, presented in the previous section, $\gone/\gzero
\sim M_\pi/m_N$. This contribution will be called leading order (\LO).

\begin{figure}[bh!]  
	\centering
  \includegraphics[scale=0.8]{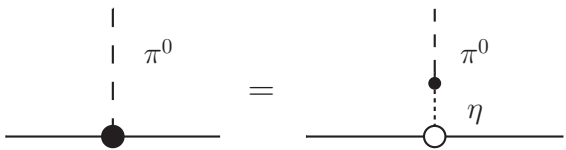}
  \caption[Isospin-odd CP-conserving pi NN vertex]
  {
   Isospin-odd  CP-conserving   $\pi NN$ vertex (black circle) induced by  $\pio$--$\eta$-mixing and
   the CP-conserving $\eta NN$ coupling (open circle).}
  \label{pioeta2}
 \end{figure}

A CP-odd pion exchange potential from a $\gzero$ coupling on one vertex and an
isospin-odd, CP-conserving coupling on the other also leads to a non-vanishing
contribution to the deuteron EDM~\cite{Lebedev_Olive,deVries:2011an}.
As long as we focus only on contributions to the deuteron EDM, the impact of
the  resulting potential is effectively a redefinition $\gone \to
\gone[1+\gzero \beta_1/(2 g_A \gone)]$~\cite{deVries:2011an}, where $\beta_1$
is the strength parameter of the isospin-odd, CP-even $\pi NN$ vertex and
$g_A$ is the axial-vector coupling constant of the nucleon. 
The Nijmegen partial-wave analysis provides $|\beta_1|\leq
10^{-2}$~\cite{nijmwegen},
which is consistent with estimating its value
from the same mechanism used in ref.\,\cite{Lebedev_Olive} and appendix~\ref{app: lebedev}, namely via 
$\eta$--$\pi^0$  mixing --- see fig.~\ref{pioeta2}. 
Thus the inclusion
of $\beta_1$ shifts $\gone$ by a few percent at most and can therefore
be neglected, given the significant uncertainty of $\gone$.

The first relativistic correction is the recoil correction to the $g_A$
vertex, given by $ -{g_A}/{(2\mn\fpi)}S\cdot(p_1+p_2)v\cdot k \tau^a $ where
$p_{1,2}$ are the nucleon momenta and $k$ is the outgoing pion momentum. The corresponding
contribution is suppressed by three orders 
relative to the one of the $g_A$ vertex 
due to the additional energy dependence (since $v =(1,\vec 0\,)$ and $k=p_1-p_2$).

To one-loop order there are a couple of diagrams as shown in  fig.\ \ref{fig:potential}.
The power counting gives for these diagrams $\gzero M_\pi/(m_N^2 F_\pi)$, where we
identified $4\pi F_\pi\sim m_N$. Thus, the loop contributions with the 
CP-violation
induced via the coupling $\gzero$ are suppressed relative
to the leading, non-vanishing contribution to the potential (proportional to
$\gone$) by one power of $M_\pi/m_N$ and therefore contribute to \NLO.
 However,
as outlined below, the spin-isospin structure of all these diagrams is such that
they do not contribute to the deuteron EDM.
At \NtwoLO\ the same topologies appear, however, with $\gzero$ replaced by
$\gone$. 
In addition, also triangle topologies of type $(d)$ with the
$\pi \pi NN$ vertex from ${\cal L}_{\pi N}^{(2)}$~\cite{BKM1995} as well
as vertex corrections (diagrams $(e)$ and $(f)$) formally appear at this
order. As shown below, besides the latter class none of the mentioned diagrams
contributes to the deuteron EDM.

On dimensional grounds CP-odd four-nucleon operators start to contribute at order
$M_\pi/m_N$ relative to the leading term. Their largest  $\theta$-term-induced contributions 
are isospin conserving  ({\it c.f.} second line of eq.~(\ref{eq: lth})). Thus, as a consequence of the
Pauli--Principle, they change the two-nucleon spin. Therefore they do not contribute to the
deuteron EDM.  However, their isospin-violating counter parts  ({\it c.f.} third line of
eq.~(\ref{eq: lth})) contribute, but have a relative suppression
of order $(M_\pi/m_N)^2$ and are therefore of \NthreeLO.
 
In summary, to the order we are working, the only contribution to the CP-odd
$NN$ potential that needs to be considered for the deuteron EDM is the
isospin-odd tree-level contribution proportional to $\gone$ and its vertex corrections.

\subsection{Power counting of the irreducible transition currents}

\begin{figure*}[t!]  
	\centering
 {\includegraphics[width=0.75\textwidth]{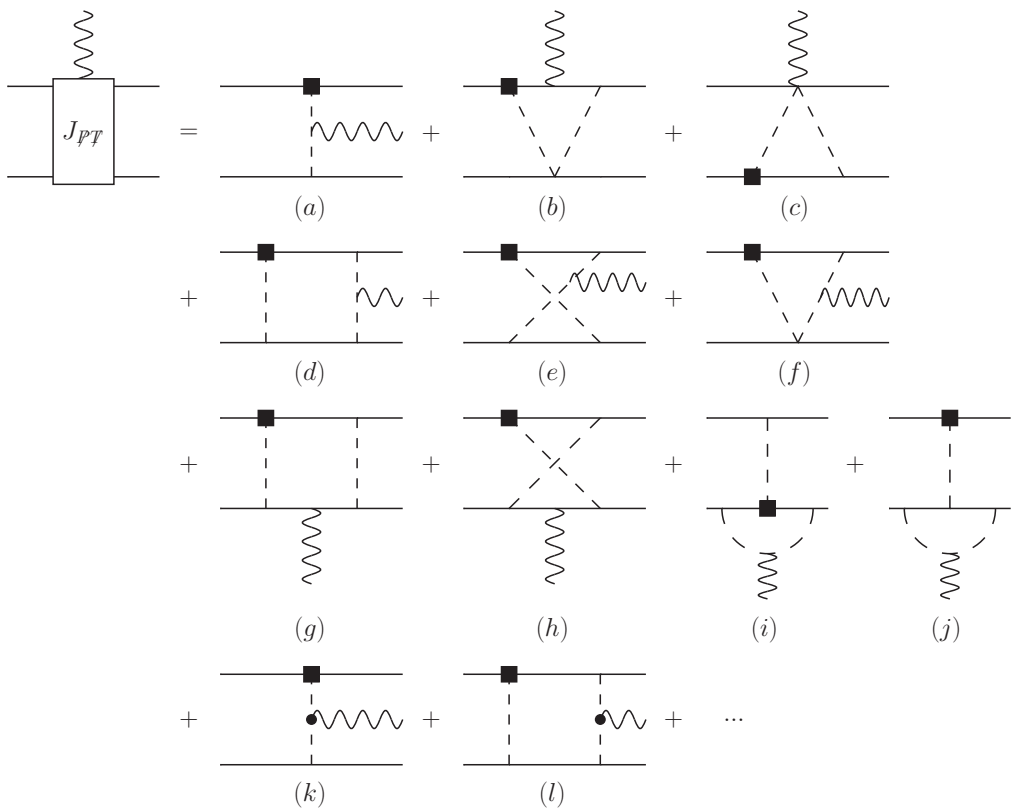}}
  \caption[CP-violating transition current]
  {
   Contributions to the CP-violating transition current: 
   (a) \NLO\  contribution, (b)-(l)
   \NtwoLO\ contributions. Solid lines denote nucleons and
dashed line denote pions. The CP-violating vertex is depicted by a black box, 
a CP-conserving, but isospin-violating vertex by a filled circle.  For each class of diagrams only one representative is shown.
    \label{fig:current}}
\end{figure*}

We now turn to the transition currents.
As explained in the beginning of this section, in order to compare the
contribution from the CP-odd $NN$ potential to that of the CP-odd
transition currents, the former needs to be multiplied by $e\,m_N/M_\pi^2$.
Thus, the {\em leading order contribution} of the total transition current is estimated to scale
as  $ \gone\,e\,\mn/(\mpi^3\fpi)\sim \gzero\,e/(\mpi^2 F_\pi)$.

The tree-level contribution, shown in fig.~\ref{fig:current}, is
formally of  \NLO, however, turns out to be of isovector character and thus does not add to the deuteron EDM.

The one-loop contributions to the irreducible transition current are estimated as
$\gzero\,e/(\mn^2\fpi)$ and are therefore of \NtwoLO.
The naive power counting of the diagram classes depicted in
fig.~\ref{fig:current}\,$(d)$ and fig.~\ref{fig:current}\,$(e)$ is slightly
more subtle due to the cancellation of one of the nucleon propagators by the
energy dependence of the $\pi\pi\gamma$-vertex. Therefore these diagrams
are part of the irreducible transition current and appear at
\NtwoLO.

Finally there are two additional structures ---
  fig.~\ref{fig:current}\,$(k)$ and $(l)$ --- that appear
since the zeroth component of the $\gamma\pi\pi$ vertex 
is proportional to the energy exchanged and thus gets sensitive
to the total neutron--proton mass difference\footnote{We would like to thank J.\,de Vries, U.\,van Kolck and
  R.\,G.\,E. Timmermans for drawing our attention to these currents.
The same effect in a
  different context is discussed in detail in ref.~\cite{CSBpn2dpi}.}, $\Mnp$.
The contributions of the diagrams of  fig.~\ref{fig:current}\,$(k)$ and $(l)$ can be estimated as
$\gzero e \Mnp/(M_\pi^3 F_\pi)$
and $\gzero e\Mnp/(\mn \mpi^2 \fpi)$, respectively.
Thus the former (latter) appears to be suppressed by $\Mnp/\mpi$ ($\Mnp/\mn$)
compared to the leading order. Based on NDA one might asign $\Mnp\sim \epsilon
\mpi^2/\mn$  such that diagram $(k)$ would appear at \NLO, while diagram $(l)$
would appear at \NtwoLO. However, as argued above the nucleon mass difference is
significantly smaller that its NDA estimate --- this observation made us asign
$\gone/\gzero\sim \mpi/\mn$, and not $(\mpi/\mn)^2$ as would follow from NDA. In full analogy we now
asign  diagram $(k)$ and diagram $(l)$ the orders \NtwoLO\ and \NthreeLO,
respectively. Therefore the former is included in our calculation while
the latter can be neglected.

In table~\ref{tab: PowerCounting} the power-counting  scales of the CP-violating
irreducible $NN$ potentials and those of the irreducible as well as of the total transition currents can be found.
This completes the discussion of the power counting. 
In the next section the
various
diagrams are discussed explicitly.

\section{EDMs from the {\boldmath$\theta$}-term} \label{sec: thetaEDM}

The computation of the two-nucleon contributions to the deuteron EDM is most efficiently performed in the Breit frame defined by 
$q= P-P'=(0,\vec{P}-\vec{P}')$ where $P$
and $P'$ denote the total four-momenta of the incoming and outgoing deuteron states
and  $q$ the momentum of the external `Coulomb-like' photon.
The electric dipole moment $d$ of the deuteron nucleus of mass $m_D$ is then defined  (in analogy to the magnetic moment case) 
by
\begin{equation}
   d=\lim_{\vec{q}\to 0}\,\frac{F_3(\vec{q}^{\,2})}{2 m_{D}}\,,
\end{equation}
where the electric dipole form factor $F_3$ is related to the P- and T-violating transition current operator $(J_{\Pv\Tv}^{\rm total})^{\mu}$ by
\begin{eqnarray} 
   &&\left \langle J=1, J'_z=\pm 1;\vec{P}\,' \left|(J_{\Pv\Tv}^{\rm total})^0 \right|J=1,J_z=\pm 1;\vec{P}\,\right\rangle \nonumber \\
 &&  \qquad  = \mp i q^3 \frac{F_3(\vec{q}^{\,2})}{2m_{D}}\,,
   \label{eq:edm}
\end{eqnarray}
where $J$ is the total angular momentum  of the deuteron and $J_z$ and $J'_{z}$ its $z$-components for the in- and out-state, respectively.

\smallskip
The total CP-violating transition current $J_{\Pv\Tv}^{\rm total}$ can be
separated into two contributions of different topology 
(see fig.~\ref{fig: total_current}): two-nucleon-reducible transition currents where the P- and
T-violation is induced by a CP-violating two\---nucleon potential on the one
hand, and irreducible CP-violating transition currents on the other.
These will now be discussed in detail.

\subsection{Contributions from the CP-odd {\boldmath $NN$} potential to the deuteron EDM} \label{sec: NN_potential}

 In order for a P- and T-violating two-nucleon potential to  contribute
 in the deuteron channel, it must induce
 ${}^3S_1$-$^3D_1\to{}^3P_1$ transitions, {\it i.e.} isospin 0 to isospin 1
 and spin 1 to spin 1 transitions since the photon-nucleon coupling is spin
 independent --- it therefore must be antisymmetric in isospin space and symmetric
 in spin space.

 Contributions to the CP-violating two-nucleon potential can be further
 separated into irreducible and reducible potentials. The latter class
 consists of a CP-violating potential and of multiple insertions of the $NN$ potential in the
 ${}^3S_1$-$^3D_1$ state and/or in the intermediate ${}^3P_1$ state, which can
 be either absorbed into the deuteron wave functions, or into the intermediate
 $NN$ interactions in the ${}^3P_1$ state and therefore do not need to
be considered separately.

The leading contribution to the CP-violating two\--nu\-cleon potential is the class of tree-level diagrams depicted 
in fig.~\ref{fig:potential}\,$(a)$. 
The tree-level potential induced by the $g_0^\theta$ vertex is given by \cite{Khriplovich:1999qr,Maekawa:2011vs}
\begin{equation}
  V_{\ref{fig:potential} (a)}(\vec{l})=i\frac{g_0^\theta g_A}{2\fpi}\frac{\vec{l}}{\vec{l}^2+\mpi^2} \cdot (\vec{\sigma}_{(1)}{-}\vec{\sigma}_{(2)})\, \vec{\tau}_{(1)}\cdot\vec{\tau}_{(2)}
    \,,
\end{equation}
where $\vec{l}$ denotes the pion momentum running from nucleon $1$ to nucleon $2$.
It is spin antisymmetric and isospin symmetric and does not induce
${}^3S_1$-$^3D_1\to{}^3P_1$ transitions~\cite{Khriplovich:1999qr,Liu_Timmermans,Maekawa:2011vs}.

The potential induced by the $g_1^{\theta}$ vertex reads 
\begin{eqnarray}
  V^{\theta\,\LO}_{\ref{fig:potential} (a)}(\vec{l})
  &=&i\frac{g_1^{\theta}g_A}{4\fpi}\frac{\vec{l}}{ \vec{l}^2+\mpi^2} \cdot \Bigl[(\vec{\sigma}_{(1)}{+}\vec{\sigma}_{(2)})
  (\tau_{(1)}^3-\tau_{(2)}^3)\nonumber \\
  && \qquad\qquad\qquad +(\vec{\sigma}_{(1)}
{-}\vec{\sigma}_{(2)})(\tau_{(1)}^3{+}\tau_{(2)}^3)\Bigr]
\end{eqnarray}
with $\vec l$ as above. It is the same as in \cite{Maekawa:2011vs,Khriplovich:1999qr,Liu_Timmermans} with $g_1^{\theta}$ replaced by $g_1$. This potential-operator has a spin-symmetric and 
isospin-antisymmetric component and thus con\-tribu\-tes to the transition current in the deuteron channel. 
In order to evaluate its contribution to the EDM of the deuteron we resort to the parametrization of the deuteron wave function of \cite{Machleidt:2000ge} with a
${}^3D_1$-state probability of $4.8\%$. In order to include the $NN$ interactions in the intermediate ${}^3P_1$-state we use the separable rank-2
representation of the Paris nucleon-nucleon potential of ref.~\cite{Haidenbauer:1984dz} (PEST). The resulting contributions to the deuteron EDM listed
in table~\ref{tab: reslo} are in agreement with the results for $g_1$ of  ref.~\cite{Liu_Timmermans} using the Argonne $v_{18}$ potential, of ref.~\cite{Liu_Timmermans,Afnan:2010xd,Gibson:2012zz}
using the Reid93 potential, and of ref.~\cite{Khriplovich:1999qr,deVries:2011re}, where the deuteron wave function has been used 
in the Zero-Range-Approximation (ZRA). The ${}^3D_1$-admixture is found to enhance the deuteron EDM by about 20\,\%, whereas the interaction in the intermediate  ${}^3P_1$-state reduces the contribution by
about  the same amount.

\begin{table*}[tb]
   \caption[Results of the tree-level calculation]
                  {Leading order contributions to the deuteron EDM  from the $\gone$ vertex without ($d^{\theta}_{PW}$, PW: plane wave) 
                   and with ($d^{\theta}_{MS}$, MS: multiple scattering)
                   intermediate ${}^3P_1$-interactions and the total leading order contribution 
                   $d^{\theta}_{\text{\LO}}$ in units of 
                   $(g_1^{\theta}/g_0^{\theta})G_{\pi}^{0}\,e\,{\rm fm}$ with
                   $G_{\pi}^{0}=g_0^{\theta}g_Am_N/\fpi$ -- calculated in Zero-Range-Approximation (ZRA),
                   with the Argonne $v_{18}$\,\cite{Argonne_v18}, Reid93\,\cite{Reid93} and CD-Bonn\,\cite{Machleidt:2000ge} potentials.                 
                    \label{tab: reslo}                  
                   }
  \centering
  \begin{tabular}{l  c c  c  c c }
\hline\noalign{\smallskip}
 &Potential					&${}^3D_1$-adm.	&$d^{\theta}_{PW}$
 &$d^{\theta}_{MS}$
 & $d^{\theta}_{\LO}$
 \\
\noalign{\smallskip}\hline\noalign{\smallskip}
 \cite{Khriplovich:1999qr,deVries:2011re} &ZRA	& \ \ ---		&$ -1.8\cdot10^{-2}$		
 &\ \ ---			&
  $-1.8\cdot10^{-2}$	
 \\
\cite{Liu_Timmermans,Timmermans_private}&A\,$v_{18}$		&$5.76\%$	&	
 	& &
  $-1.43\cdot10^{-2}$	
  \\
  \cite{Liu_Timmermans,Timmermans_private}&Reid93		&$5.7\%$	&	
 	& &
  $-1.45\cdot10^{-2}$	
  \\
 \cite{Afnan:2010xd,Gibson:2012zz}&Reid93			&$5.7\%$	&$-1.93\cdot10^{-2}$	
 	&$0.40\cdot10^{-2}$	&
  $-1.53\cdot10^{-2}$	
\\
 This work 
 &CD-Bonn						&$4.8\%$	&$-1.95\cdot10^{-2}$		&$0.44\cdot10^{-2}$	&
 $-1.52\cdot10^{-2}$	
 \\
\noalign{\smallskip}\hline
 \end{tabular}
\end{table*}

Loops formally start to contribute at \NLO.
The reducible component of the box potential of
fig.~\ref{fig:potential}\,$(b)$ constitutes a static one-pion exchange and is already accounted for either by the deuteron
wave functions or by the interaction in the intermediate ${}^3P_1$-state. Its irreducible component 
may be obtained by shifting the pole of one of the nucleon propagators into the half plane of the pole of the other nucleon
propagator, as outlined in \cite{Kaiser:1997mw,Kaiser:1999jg,Zhu:2004vw}:
$
i/(-v\cdot p_i+i\epsilon)\rightarrow{-i}/({v\cdot p_i+i\epsilon}).$
For the sum of the irreducible part of the box potential of fig.~\ref{fig:potential}\,$(b)$ 
and  the crossed-box potential of fig.~\ref{fig:potential}\,$(c)$, 
one finds in dimensional regularization in $d$ space--time dimensions
\begin{eqnarray}
  V_{\ref{fig:potential} (b{+}c)}^{\theta\,\NLO}(\vec{l})
  &=&-i\frac{\gzero g_A^3}{16\pi^2\fpi^3}\frac{1+\frac{3}{2}\xi}{\sqrt{\xi(1+\xi)}}\ln\!\left(\frac{\sqrt{1+\xi}
  +\sqrt{\xi}}  {\sqrt{1+\xi}-\sqrt{\xi}}\right) \nonumber\\
&& \mbox{} \times  \vec{\tau } _ { (1)}\cdot\vec { \tau }_{(2)}
  ({\vec\sigma}_{(1)}-{\vec \sigma}_{(2)})\cdot \vec{l}
\end{eqnarray}
with $\xi={\vec l}^2/(4\mpi^2)$. Note that the divergence has been absorbed by a redefinition of the four-nucleon coupling 
constant $C_2^0$ (the scale $\mu$ is introduced in dimensional regularization)
\begin{equation}
   {C_2^0}\rightarrow {C_2^0}-\, \frac{\gzero g_A^3}{\fpi^3}\left[6L-\frac{3}{16\pi^2}\left(\ln\!\left(\frac{\mu^2}{\mpi^2}\right){-}1\right)
   -\frac{2}{16\pi^2}\right]
\end{equation}
with
\begin{equation}
  L = \frac{\mu^{d-4}}{16\pi^2} \left\{ \frac{ 1}{d-4} +\frac{1}{2} \bigl[\gamma_E - 1 - \ln(4\pi) \bigr] \right\} \,
\end{equation}
where $\gamma_E=0.577215\cdots$ is the Euler--Mascheroni constant.

The triangular potential of
fig.~\ref{fig:potential}\,$(d)$ gives
\begin{eqnarray}
      V_{\ref{fig:potential} (d)}^{\theta\,\NLO}(\vec{l})&=& i\frac{\gzero g_A}{32 \pi^2\fpi^3}\sqrt{\frac{1+\xi}{\xi}}
    \ln\!\left(\frac{\sqrt{1+\xi}+\sqrt{\xi}}{\sqrt{1+\xi}-\sqrt{\xi}}\right)\nonumber \\
  && \mbox{} \times    \vec{\tau}_{(1)}\cdot\vec { \tau }_{(2)}
     ({\vec\sigma}_{(1)}-{\vec \sigma}_{(2)})\cdot \vec{l}
\end{eqnarray}
where the divergence has been absorbed by a further redefinition of $C_2^0$:
\begin{equation}
  {C_2^0}\rightarrow {C_2^0}-\,\frac{\gzero g_A}{\fpi^3}\left[-2L+\frac{1}{16\pi^2}\left(\ln\!\left(\frac{\mu^2}{\mpi^2}\right){-}1\right)+\frac{2}{16\pi^2}\right].
\end{equation}
These results reproduce those of ref.~\cite{Maekawa:2011vs}. 
Note that all $g_0^{\theta}$ potential-operators up to one loop as well as the four-nucleon-vertex operators are
isospin symmetric and spin antisymmetric and therefore vanish in the deuteron
channel.

At \NtwoLO\ there are  the same topologies as just discussed,
however, with the $\gzero$ vertex replaced by its isospin-violating counter
part $\gone$.
The triangular-potential
operator fig.~\ref{fig:potential}\,$(d)$ vanishes at the considered order. The
class of the crossed-box-potential diagrams of fig.~\ref{fig:potential}\,$(c)$
gives

\begin{eqnarray}
V_{\ref{fig:potential} (c)}^{\theta\,\textrm{\NtwoLO}}(\vec{l})&=&-i\frac{\gone g_A^3}{8\fpi^3}\Bigg\{\frac{1}{16\pi^2}\frac{1{+}\frac{3}{2}\xi}
{\sqrt{\xi(1{+}\xi)}}\ln\!\left(\frac{\sqrt{1{+}\xi}+\sqrt{\xi}}{\sqrt{
1{+}\xi } -\sqrt { \xi } }
\right) \nonumber \\
&&  +\left[3L-\frac{3}{2}\frac{1}{16\pi^2}\left(\ln\!\left(\frac{\mu^2}{\mpi^2}\right){-}1\right){-}\frac{1}{16\pi^2}\right]\Bigg\} \nonumber \\
& &\quad \mbox{}\times\Bigl[(\tau^3_{(1)}-\tau^3_{(2)})    ({\vec\sigma}_{(1)}+{\vec\sigma}_{(2)})\nonumber \\
&& \qquad\ \mbox{}+(\tau^3_{(1)}+\tau^3_{(2)}){(\vec\sigma}_{(1)}-{\vec\sigma}_{(2)})   \Bigr] \cdot  {\vec{l}}\,.
\label{crossedboxg1}
\end{eqnarray}
Resorting again to the method presented in \cite{Kaiser:1997mw,Kaiser:1999jg,Zhu:2004vw} to isolate the irreducible component of the box potential-operator
fig.~\ref{fig:potential}\,$(b)$, the  latter is found to be the negative of eq.~(\ref{crossedboxg1}) and to
cancel
the crossed-box-potential-operator 
fig.~\ref{fig:potential}\,$(c)$. Therefore, contributions to the total CP-violating transition current induced by the
CP-vio\-lating two-nucleon one-loop potential are absent to \NtwoLO\ ---  not only in the deuteron channel.

The only non-vanishing \NtwoLO\ contributions are thus the vertex corrections
shown in diagrams  \ref{fig:potential}\,$(e)$ and $(f)$. The vertex correction on the CP-conserving
vertex is readily accounted for,  since we use the physical $\pi NN$ coupling
constant in our calculations. The situation is somewhat different for diagram
\ref{fig:potential}\,$(e)$, where the physical value of the coupling constant is not known, but was  
calculated/estimated in sect.~\ref{sec:piNNCPv}.
Since $\gzero$ only appears at the one-loop level in the case of the deuteron EDM, 
we only need to consider $\gone$ here.
The quoted uncertainty for
$\gone$ is of the order of 50\%. On the other hand, the corresponding
correction for the CP-conserving $\pi NN$ coupling constant, the so called
Goldberger--Treiman discrepancy, is very small~\cite{veroniquereview}, such that we may safely assume
that the uncertainty given for $\gone$ is sufficiently large such that it includes
vertex corrections.

 Thus, the only piece of the $NN$ potential that is CP odd {\em and}
 contributes to the deuteron EDM is the tree-level diagram depicted in
 fig.~\ref{fig:potential}\,$(a)$, with the $g_1^\theta$ coupling employed in the
 CP-odd $\pi NN$ vertex: it is the  \LO\ potential.

\subsection{Contributions from the CP-odd irreducible {\boldmath $NN$} transition current} \label{sec: Transition_Current}

 In order for an irreducible transition current not to vanish in the deuteron
 channel, it has to induce ${}^3S_1$-$^3D_1\to {}^3S_1$-$^3D_1$ (isospin 0 to
 isospin 0 and spin 1 to spin 1) transitions. It therefore needs to be an
 isoscalar operator, symmetric in spin space.  Therefore the tree-level
 transition currents --- {\it c.f.} fig.~\ref{fig:current}\,$(a)$ --- that are
 all isovector in character, do not contribute to the deuteron EDM.  The
 relevant CP-odd irreducible one-loop $NN$ current operators are listed in
 fig.~\ref{fig:current}\,$(b)$-$(j)$.  Diagrams involving CP-even
 $NN\pi\gamma$-vertices have been neglected here since, to the order we are working, they do
 not yield EDM contributions: according to eq.~(\ref{eq:edm}) EDM
 contributions are extracted from the $0$th-component of matrix elements
 of transition currents. The leading order, CP-even $NN\pi\gamma$ vertex $ i
 e(g_A/\fpi)\,\varepsilon\cdot S\epsilon^{a3b}\tau^b $ (see appendix A of
 \cite{BKM1995} where $\gamma$ is the ``Coulomb photon'', $\varepsilon=(1,\vec
 0)$) does not have a non-vanishing $0$th-component for $S=(0,{\vec\sigma}/2)$.

The diagram classes depicted in 
fig.~\ref{fig:current}\,$(g)$ and fig.~\ref{fig:current}\,$(h)$ are of order $\gzero\,e/(\mn^2\fpi)$ and thus
 \NtwoLO. For a photon coupling to nucleon 2 the two-nucleon-irreducible component of diagram 
fig.~\ref{fig:current}\,$(g)$ and diagram fig.~\ref{fig:current}\,$(h)$
give
\begin{eqnarray} 
&&\left(J^{\theta\,\textrm{\NtwoLO}}_{\ref{fig:current}(g{+}h)}\right)^\mu
= i\frac{eg_0^{\theta}g_A^3}{128\pi\fpi^3\mpi}\left[\frac{1}{1+\xi}+\frac{2}{\sqrt{\xi}}\arctan\sqrt{\xi}\right]  v^\mu\nonumber \\
&& \mbox{}\times
(\vec{\tau}_{(1)}\cdot\vec{\tau}_{(2)}{-}\tau_{(2)}^3)(\vec{\sigma}_{(1)}{-}\vec{\sigma}_{(2)})\cdot(\vec{p}_2'-\vec{p}_2+\vec{q})\quad\
\label{eq: boxcurrent1}
\end{eqnarray}
with $\xi= \left |\vec{p}_2'-\vec{p}_2+\vec{q}\, \right|^2/(4\mpi^2)$ in terms 
of the initial (final) momentum $p_i (p_i' )$ of nucleon $i$ and the momentum of the out-going photon $q$.
 Although the operator (\ref{eq: boxcurrent1}) contains an 
isospin-symmetric component, it is spin-antisymmetric and vanishes in the deuteron channel.

The diagram classes depicted in 
fig.~\ref{fig:current}\,$(d)$ and $(e)$
 vanish in the deuteron channel, since they are isovectors.

In addition there are diagrams at \NtwoLO\ where the photon couples to a
vertex correction (fig.~\ref{fig:current}\,$(i)$ and $(j)$); however, terms
that contain the $\gzero$ vertex turn out to be isovectors and thus do not
contribute to the deuteron channel, and those that contain $\gone$ start
to contribute only at \NthreeLO.

The triangular diagrams depicted in 
fig.~\ref{fig:current}\,$(b)$, fig.~\ref{fig:current}\,$(c)$ and fig.~\ref{fig:current}\,$(f)$ are all of 
order \NtwoLO. 
Diagrams of the types of 
fig.~\ref{fig:current}\,$(c)$ and fig.~\ref{fig:current}\,$(f)$  vanish in the deuteron channel which can  be readily seen from their isospin
components: diagram 
fig.~\ref{fig:current}\,$(c)$ is proportional to $\tau_{(2)}^3$ (photon coupling to nucleon 2\,) and diagram 
fig.~\ref{fig:current}\,$(f)$ is proportional to $2\tau_{(2)}+i(\vec{\tau}_{(1)}\times\vec{\tau}_{(2)})$.
A class of currents that has a spin- and isospin-symmetric component
 is  depicted in
fig.~\ref{fig:current}\,$(b)$:
\begin{eqnarray}
 \left( J^{\theta\,\textrm{\NtwoLO}}_{\ref{fig:current}(b)}\right)^\mu&=&
 i \frac{ eg_0^{\theta}g_A}{4\fpi^3}\,v^{\mu}
 \,\left(\vec{\tau}_{(1)}\cdot\vec{\tau}_{(2)}
 -\tau^3_{(2)}\right) \nonumber \\
 &\times&   \left(I(p_1{-}p_1')\,(\vec{p}_1'{-}{\vec p}_1) \cdot {\vec \sigma}_{(2)}  + (1\leftrightarrow 2)\right)
 \ \ \ \ \ \ \label{eq:master}
\end{eqnarray}
with 
$I(l)$ = $-\arctan(|\vec{l}\,|/(2\mpi))/{(8\pi|\vec{l}\,|)}$~\cite{BKM1995}.
Resorting to the CD-Bonn wave function of the deuteron as used above,
the resulting $g_0^{\theta}$-contribution to the deuteron EDM for the ${}^3S_1$ state and ${}^3D_1$
admixture  is found to be
\begin{equation}
  d_{ \ref{fig:current}(b) }^{\theta} = 
  \underbrace{-2.00\cdot 10^{-4}
            \,\times G_{\pi}^{0}\, e\, {\rm fm}}_{{}^3S_1}
 -\underbrace{0.53\cdot 10^{-4}
            \,\times G_{\pi}^{0} \,e\,{\rm fm}}_{{}^3D_1\text{-adm. }}
            \label{dN2LO}
\end{equation}
where $G_{\pi}^{0} :=g_0^{\theta}g_Am_N/\fpi$.

The  class of diagrams depicted in 
fig.~\ref{fig:current}\,$( k )$, see ref.\,\cite{deVries:2011an},
gives
\begin{eqnarray}
&&\left( J^{\theta\,\textrm{\NtwoLO}}_{\ref{fig:current}(k)}\right)^0
= -i\frac{e\gzero g_A \Mnp}{\fpi}(\vec{\tau}_{(1)}\cdot\vec{\tau}_{(2)}-\tau_{(1)}^3\tau_{(2)}^3)\nonumber\\
&&
\mbox{}\times \frac{{\vec\sigma}_{(1)}\cdot({\vec p}_1{-}{\vec p}_1')+ {\vec\sigma}_{(2)}\cdot ({\vec p}_2-{\vec p}_2') }  
  {\left[(\vec{p}_1\!-\!\vec{p} _1')^2 +\mpi^2\right]
    \left[(\vec{p}_2{-}\vec{p}_2')^2 +\mpi^2\right]}\, .
\end{eqnarray}
 The explicit
evaluation of the EDM contribution of fig.~\ref{fig:current}\,$(k)$ yields 
$0.31\cdot10^{-4}\times G_{\pi}^0\, e$\,fm, 
which justifies
the classification as \NtwoLO.

The absence of {\em both} --- divergences and (undetermined) counter-terms up to 
\NtwoLO ---
ensures the predictive power of the two-nucleon contributions to the
deuteron EDM that is induced by the $\theta$-term. 
Together with the $g_1^{\theta}$ contribution the total 
two-nucleon contribution to the EDM of the deuteron
induced by the $\theta$-term is then given by:
\begin{eqnarray}
  \label{eq:dtheta_res}
   d^{\theta} &=&d^{\theta}_{\text{\LO}}\,+\,d^{\theta}_{\text{\NtwoLO}}\nonumber \\
 &=&  \left[\left(-15.2\cdot \frac{g_1^{\theta}}{g_0^{\theta}}
   -0.22\right)\pm 0.03\right] {\times} 10^{-3}\, G_{\pi}^{0}\, e\, {\rm fm}\,,  \ \ \ \ \ 
\end{eqnarray}
where the uncertainty estimates the higher order contributions not included as
given by the 
 power counting.
Alternatively we may express the result directly in terms of $\bar \theta$, the strength of
the QCD $\theta$-term, and write
\begin{eqnarray}
    d^{\theta} &=&d^{\theta}_{\text{\LO}}\,+\,d^{\theta}_{\text{\NtwoLO}}\nonumber \\
    &=&
      -\big(( 5.9\pm 3.9) - (0.5\pm 0.2)\big) \times 10^{-4}  \,\btheta  \,e\,{\rm  fm}\,,
 \label{eq:dtheta_final}
\end{eqnarray}
where the uncertainties now contain, in addition to the one
given in eq.~(\ref{eq:dtheta_res}), also the uncertainties in the coupling
constants
$\gzero$ and $\gone$.
Therefore the final result is completely dominated by the contribution from
the CP- and isospin-violating tree-level potential proportional to $\gone$.

\section{Summary and conclusions} \label{sec: Conclusions}

As already stated in the introduction, the established relation between the
QCD $\theta$-term and the CP-odd $\pi NN$ coupling constant is not sufficient
to predict the size of the electric dipole moment of a {\em single} nucleon
(neutron or proton) with the help of effective field theory, since the
calculable one-loop contributions are of the same order as undetermined
counter terms.  However, this unpleasant feature is not present for the
two-nucleon contributions of the deuteron and other light nuclei, which
contribute already at tree-level order --- unaffected by any counter terms ---
and which can be derived --- admittedly with a large uncertainty ---
up-to-and-including the order \NtwoLO, see eqs.~ (\ref{eq:dtheta_res}) and
(\ref{eq:dtheta_final}) at the end of sect.~\ref{sec: Transition_Current}.
The \NtwoLO\ contributions of these results are (up to vertex corrections
discussed in sect.~\ref{power}) solely governed by the irreducible transition
currents. The latter  include loops which for the first time have been calculated in
the present work. Note that any contribution with unknown coefficients can
only show up at \NthreeLO.

The dominant part of the deuteron's two-nucleon EDM from the QCD $\theta$-term
resulted from an isospin-violating, CP-odd $\pi NN$ coupling constant, $g_1$.  
The isospin-violation of this coupling can be estimated from the strong contribution 
to the pion mass-square splitting $\Mpistr/(\mpi^2\epsilon)$.
Although this ratio gives a small number, its contribution to $\gone$ gets enhanced
 by  the relatively large pion-nucleon sigma term. 
Nominally, $\gone$ should be suppressed by two orders  relative to its
isospin-conserving counter part,
$\gzero$.
However, the latter is governed by the
 strong part of the neutron-proton mass splitting and therefore is found
to be exceptionably small. Thus
the isospin-violating coupling $\gone$ --- as already observed by
Lebedev \textit{et al.}\,\cite{Lebedev_Olive} --- is effectively only suppressed by one
power in the counting.

This is important since the one-pion exchange with one $\gzero$ vertex cannot
contribute to the two-nucleon part of the deuteron EDM because of
isospin selection. This was summarized in the folklore that the deuteron would
be blind to the two-nucleon contributions generated by the $\theta$-term.
This folklore, however, should be abandoned. A measurement of  a non-vanishing
neutron,  a non-vanishing proton and  a non-vanishing deuteron EDM would suffice to determine the strength of
the QCD $\theta$-term, $\btheta$, from data.  
Note that the two-nucleon part
of the deuteron EDM given in (\ref{eq:dtheta_final}) is in fact of the same
magnitude and therefore comparable in size with the non-analytic isovector
part of the nucleon EDM as calculated in ref.~\cite{Ottnad}, which is,
using as input the value of $\gzero$ from eq.~(\ref{gzero_form}),
\begin{equation}
 d_N^{\,\textrm{non-analyt.}} 
 = {(21\pm 9) \times 10^{-4}  \,\bar \theta \, e\, {\rm fm}\, ,} 
\end{equation}
where the uncertainty contains both the variation of the loop scale as
proposed
in ref.~\cite{Ottnad} as well as the uncertainty in $\gzero$. This number
may presumably be taken as a  scale  which  governs  the single nucleon EDMs. Note, however, that the non-analytic contribution
to the isoscalar part of the nucleon EDM is an order of magnitude smaller due
to a suppression by a factor $M_\pi/m_N$ as well as the absence of a chiral
logarithm.
Whether the proton or neutron EDM
are really of the same magnitude as the two-nucleon part of the deuteron EDM
is a question which {\em only} experiments might eventually be able to answer.  

Fact
is that, under the assumption that the electric dipole moments are driven by the CP violation that is
induced by the QCD $\theta$-term, we now can give a relation between the total
EDMs of the deuteron,  the neutron and  the proton and the calculated two-nucleon EDM part
of the deuteron:
\begin{equation}
 d_{D} = d_n + d_p    -\big(( 5.9\pm 3.9) - (0.5\pm 0.2)\big)\times 10^{-4} \,\bar \theta \, e\,{\rm fm}. 
  \label{last word}
\end{equation}
A cross-check of the so-extracted $\btheta$ value would be possible ---
still solely from data --- by a measurement of the EDM of
${}^3$He. Another strategy to test or falsify the $\btheta$ value would
involve lattice QCD calculations and just two successful EDM measurements,
namely one single-nucleon EDM, {\it i.e.} the one of the neutron or proton,
and the deuteron EDM.  If even all three of them are measured, then one could
use lattice QCD for a first test correlating the proton and neutron EDM
results in terms of the parameter $\btheta$ and to use formula (\ref{last
  word}) for an additional, orthogonal test.

If indeed the QCD $\theta$-term would have failed these tests --- either by a
direct comparison of data or by the additional involvement of lattice QCD ---
then  the following picture would emerge: 
in case
$d_D-d_n-d_p$ is sizable {\em compared to what eq.~(\ref{last word})  in combination with experimental or lattice data predicts},
then
the dimensional analysis reveals a dominance of the quark-color EDM, feeding
the coupling {proportional to $g_1$}~\footnote{Note that ref.~\cite{deVries:2011an} stated the dominance of the
quark-color mechanism already under the assumption that  $d_D-d_n-d_p$ itself is sizable.
The difference emerges since in  ref.~\cite{deVries:2011an} the relative
suppression between $\gone$ and $\gzero$ was taken from naive dimensonal
analysis that predicts a negligible contribution from the $\gone$ term.}.
On the other hand, if this difference
is very small, most probably neither the $\theta$-term nor the quark-color 
EDM is at work, but one or several of the other dimension six
CP-violating operators~\cite{deVries:2011an,deVries:2010ah}. More insight can be gained from a study of the
EDM for $^3$He.
This reasoning stresses once more the need for high-precision measurements,
not only of the neutron EDM but also of the EDMs for light ions like proton,
deuteron and $^3$He.


\begin{acknowledgement}

We would like to thank W.~Bernreuther,  E.~Epelbaum, F.-K. Guo, J.~Haidenbauer,  U.~van Kolck, B.~Kubis, E.~Mereghetti,
N.\,N.~Nikolaev, J.~Pretz, F.~Rath\-mann, R.\,G.\,E.~Timmermans and J.~de Vries  for helpful discussions and T.~L\"ahde also
for advice on the numerical analysis.
This work is supported in part by the DFG and the NSFC through
funds provided to the Sino-German CRC~110 ``Symmetries and
the Emergence of Structure in QCD'', and by the European
Com\-muni\-ty-Research  Infrastructure  Integrating  Activity  ``Study of
Strong\-ly Interacting Matter'' (acronym Hadron\-Physics3).
The numerical calculations were partly performed on the supercomputer cluster of
the JSC, J\"ulich, Germany.
\end{acknowledgement}
\setcounter{equation}{0}
\begin{appendix}
\renewcommand{\theequation}{\thesection.\arabic{equation}} 

\section{Selection of the ground state}
\label{app: vacuum}

As pointed out in \cite{Dashen:1970et,Baluni} the presence of a term in the Lagrangian 
which explicitly breaks the SU(2)$\times$ SU(2) symmetry imposes a
constraint on the selection of the ground state, such that the SU(2) subgroup
to which SU(2)$\times $SU(2) is broken  is uniquely specified. This implies
especially the absence of pion tadpoles.
Therefore, the incorporation of CP-viola\-ting and chiral-symmetry-breaking
terms into the Lagran\-gian requires, in general, an adjustment of the vacuum, {\it
  i.e.} an axial transformation $A(=R=L^{\dagger})$,
\begin{equation}
  U\,\mapsto\,AUA\quad N\mapsto K(A^\dagger,A,U)N
  \label{eq: Arot}
\end{equation}
with $U=u^2$ (see {\it e.g.} \cite{BKM1995}). In the representation which we are using, the $\theta$-term is related to
the isospin-breaking mass term by an axial rotation that contains $\tau_3$
only,  $A=\exp(i\alpha\tau_3/2)$ --- {\it c.f.} the discussion at the beginning of sect.~\ref{sec:piNNCPv}. 
Since CP violation is a small perturbation, it will slightly shift
the ground state $U_0=\unitmatrix$ according to $U_0\mapsto AU_0A$. The
rotation angle $\alpha$ is determined by minimizing the potential $V$ in the
vicinity of the ground state $U_0$: 
\begin{equation}
 \partial V[U=AU_0A]/\partial\alpha=0\,.
\end{equation}
The term  $ \fpi^2 \langle\chi_+\rangle/4$
belonging to the second-order Lagrangian in the pion sector  (see  $\mathcal{L}_1$ in ref.\,\cite{Gasser:1983yg}) would, by itself,  not induce a vacuum 
shift ({\it i.e.} $\alpha(\btheta)=0$), since
the pseudoscalar source $p$ in eq.~(\ref{eq: chi_pm}) is purely isoscalar.
Thus leading order tadpoles are avoided.
However, the $l_7$ term
of the subleading fourth-order Lagrangian in the pion sector, see $\mathcal{L}_2$ and eq.\,(5.5) in ref.\,\cite{Gasser:1983yg}, does give rise to
a ($\pi_3$) tadpole term: 
\begin{equation}
    -\frac{l_7}{16}\langle \chi_-\rangle^2
    =    - l_7 (1-\epsilon^2) \epsilon \mpi^4 \,\btheta \,\frac{\pi_3}{\fpi}\left(1 -\frac{2\,\pi^2}{3\,\fpi^2}  \right)+\cdots.
\end{equation}
Therefore, 
this tadpole contribution has to be canceled by a perturbative shift of the leading-order term
 $ \fpi^2 \langle\chi_+\rangle/4$ that is
induced by an axial rotation of the ground state  by the  small angle
\begin{equation}
\alpha'(\btheta)=-l_7 (1-\epsilon^2)\epsilon\frac{\mpi^2}{\fpi^2}\btheta+\mathcal{O}(\bar{\theta}^2) \,.
\end{equation}
Note that the $\mathcal{L}_2$ terms proportional to $l_1,l_2,l_5,l_6$ as well as to the
so-called high-energy constants, see ref.\,\cite{Gasser:1983yg},  are invariant under the rotation $A$. Furthermore,  the
$\mathcal{L}_2$ term proportional to $l_4$ does not contribute either,
since here the sole external current is the
electromagnetic field and since $[\chi,Q]=0$ (Q: quark charge matrix). 
Thus there remain only  higher-order contributions which are  
generated by  the $l_3$ and $l_7$ terms of $\mathcal{L}_2$. These contributions scale as the sixth-order order terms of the pion-sector Lagrangian, {\it i.e.}
as  $\mathcal{L}_3$ in the notation of ref.\,\cite{Gasser:1983yg}, and can be neglected here.

Note, however,
that the redefinition of the ground state also induces new structures into the
{\em pion-nucleon} Lagran\-gian  \cite{Mereghetti:2010tp}, namely
\begin{eqnarray}
    c_1\langle \chi_+\rangle N^\dagger N 
       &\rightarrow& -4\alpha'(\btheta)c_1 \mpi^2 \frac{\pi_3}{\fpi}\left(1-\frac{\pi^2}{6\,\fpi^2}\right) N^\dagger N+\cdots,
               \nonumber\\
   c_5 N^\dagger \hat{\chi}_+N 
    &\rightarrow& -2\alpha'(\btheta)c_5 \epsilon \mpi^2  N^\dagger \!\left(\frac{\vec{\pi}\cdot\vec{\tau}}{\fpi}
           -\frac{(1{-}\epsilon^2)\bar\theta\,\tau_3}{2\epsilon}\!\right) N\nn\
 \\ && \qquad\mbox{}+\cdots. 
\label{eq: new structures}
\end{eqnarray}
The terms proportional to 
$c_2$, $c_3$, $c_4$, $c_6$ and $c_7$ in the pion-nucleon Lagrangian \cite{FMS98}
 are invariant under the axial rotation $A$ when the electromagnetic field is the sole external current.
The $c_5\tau_3$ term  is proportional to
$\mathcal{O}(\bar{\theta}^2)$ and can be disregarded. While the remaining $c_5$ term in
(\ref{eq: new structures}) provides a correction to the value of $\gzero$, the
term proportional to $c_1$ is a new structure: a $\gone$-vertex which is driven by the low
energy constant $l_7$.  The latter is related to the strong-interaction part of  the pion mass-square shift
$\Mpistr$ by \cite{Gasser:1983yg}:
\begin{eqnarray}
   \Mpistr
    &:=&
               \left(M_{\pi^+}^2-M_{\pi^0}^2\right)\big |_{\rm strong} \nn\\
    & =&
              2(m_u-m_d)^2B^2l_7/\fpi^2\,+\cdots\ \nn \\
    &\approx& 
              (7\,\mev)^2 \approx 2 M_\pi \cdot 0.18 \mev\, .
\end{eqnarray}
This leads to  eq.\,(\ref{eq: gone first}), {\it i.e.}
\begin{equation}
    \gone= \frac{2\,c_1\,\Mpistr\,(1-\epsilon^2)}{\fpi\,\epsilon}\bar{\theta}\,,
    \label{eq: gone vacuum}
\end{equation}
which agrees with the corresponding term in eq.\,(113) of \cite{Mereghetti:2010tp}. 
Finally,
the correction to $\gzero$ is given by
\begin{equation}
        \delta\gzero
        = \frac{\Mstr\,(1-\epsilon^2)}{4\fpi\,\epsilon}\,\bar{\theta}\,\frac{ \Mpistr }{\mpi^2}
        = \gzero\,\frac{ \Mpistr }{\mpi^2}\,,
\end{equation}
reproducing the corresponding term in eq.\,(113) in \cite{Mereghetti:2010tp}.

\setcounter{equation}{0}
\renewcommand{\theequation}{\thesection.\arabic{equation}}

\section{An update of the derivation of Lebedev \textit{et al.}\,\cite{Lebedev_Olive}}
\label{app: lebedev}

\begin{table*}[tb!]
   \caption[SU(3) values]
                  {The value of $\gzero$, $\gone$, and the ratio $\gone/\gzero$ predicted from eqs.~(\ref{gzero_su3}) 
                  and (\ref{gone_su3}) with (i) the original  SU(3) parameters $b_D$ and $b_F$ of ref.~\cite{BKM1995}, 
                  with (ii) the alternative set of  parameters based on eqs.~(\ref{eq:bdbfiso}) and (\ref{eq:bfnew}),
                  (iii) in the case that $b_D+b_F$ of (i) are replaced by $c_5$ of eq.~(\ref{eq: c5_value}). 
                  The listed uncertainties do not contain systematical SU(3) errors. \label{tab: su3}
                   }
  \centering
  \smallskip
  \begin{tabular}{c l c c c}
  \hline\noalign{\smallskip}
                           &           & $\gzero$ [$\btheta$]  & $\gone$   [$\btheta$]   & $\gone/\gzero$ \\
\noalign{\smallskip}\hline\noalign{\smallskip}
  (i) &$b_D$ \& $b_F$ from  \cite{BKM1995}    &  $-0.026 \pm 0.002$ & $ 0.00092 \pm 0.00017$ & $-0.036 \pm 0.007$ \\
  (ii)& $b_D$ \& $b_F$  alternative        &   $-0.023 \pm 0.005$ & $ 0.00088 \pm 0.00016$ & $-0.038\pm 0.011$ \\
  (iii)& $b_D\!+b_F\to c_5$ &   $-0.018 \pm 0.007$ & $ 0.00092  \pm 0.00017$  & $-0.051\pm  0.022$\\
 \noalign{\smallskip}\hline
\end{tabular}
 \end{table*}

In addition to the usual parametrization of the $\theta$-term-induced isospin-conserving  and CP-violating $\pi NN $ coupling
\begin{equation}
  \gzero = \frac{\mstar \btheta}{F_\pi} \langle N | \bar u u - \bar d d | N\rangle  \label{gzero}
\end{equation}  
the authors of ref.~\cite{Lebedev_Olive} introduced --- via the $\pio$--$\eta$ mixing\footnote{Actually, via the $\pio$--$\eta_8$ mixing.
For consistency,  we replaced here their $\pio$-$\eta$ mixing angle  by the customary one
of chiral perturbation theory \,\cite{GassLeut_eta}  --- note the explicit  $\hat m$ subtraction in the denominator.} --- the isospin-breaking counter  part
\begin{equation}
   \gone =  \frac{\mstar \btheta}{F_\pi} \,\frac{\sqrt{3}(m_d-m_u)}{4 (m_s-\hat m)}\,
   \frac{1}{\sqrt{3}} \langle N | \bar u u + \bar d d  - 2 \bar s s| N\rangle \  . \label{gone}
\end{equation}
This is an alternative derivation of  the vacuum-alignment result (\ref{eq: gone vacuum}) for $\gone$, 
discussed in appendix~\ref{app: vacuum}, because the $l_7$ coefficient
of the fourth-order Lagrangian effectively summarizes the $\pio$--$\eta$ mixing by the quark-mass dependent shift to the pion-mass-square
$\Mpistr$.

Inserting the strong-interaction contribution to the neu\-tron-proton mass difference
$ (m_u-m_d) \langle N | \bar u u - \bar d d | N\rangle =  \Mstr $ and utilizing  the parameter $\epsilon$  as defined at the beginning
of sect.~\ref{sec:piNNCPv},
we derive (\ref{gzero_form}) again:
\[
 \gzero = \frac{\Mstr ( 1- \epsilon^2)}{4 F_\pi \epsilon} \,\bar\theta \,.
\]
 
Similarly, starting now from  eq.~(\ref{gone}), we get 
\begin{equation}
 \gone = 
        \frac{- \btheta}{8F_\pi} \, (1-\epsilon^2)\, \epsilon\, \frac{M_\pi^2}
 {M_K^2 \!- \! M_\pi^2}\, \hat m \langle N | \bar u u \!+\! \bar d d  \!-\! 2 \bar s s| N\rangle
  \end{equation}
 with
 $M_\pi^2 = 2B \hat m+ {\cal O}({\cal M}^2)$ and 
 $M_K^2 = B (m_s + \hat m) + {\cal O}({\cal M}^2)$ for the square of the pion and kaon mass, respectively, where here
 ${\cal M}$ is the quark mass matrix for three light flavors.
According to refs.~\cite{Gasser:1982ap,Borasoy_Meissner} we have
$
\hat m \langle N | \bar u u \!+\! \bar d d  \!-\! 2 \bar s s| N\rangle = \hat m \langle N | \bar u u \!+\! \bar d d |N \rangle
(1-y)
$
with
$ \sigma_{\pi N} \equiv\sigma_{\pi N}(0) =\ \hat m \langle p | \bar u u + \bar d d | p\rangle$ and
$y \equiv  {2 \langle p | \bar s s | p\rangle}/{\langle p| \bar u u + \bar d d |p \rangle}$,
where $|p\rangle$ denotes here the proton state. The final result is therefore
\begin{equation}
\gone
= - \frac{\btheta}{8F_\pi} \, (1-\epsilon^2)\, \epsilon\, \frac{M_\pi^2}
 {M_K^2 \!- \!  M_\pi^2}\, \sigma_{\pi N} (1-y)\,. 
 \label{gone_form}
\end{equation}
Inserting
$\Mstr  = (2.6 \pm 0.5)\,{\rm MeV}$ from ref.~\cite{Walker-Loud}, 
$F_\pi$  =  $92.2\,{\rm MeV}$,
and the $\overline{MS}$ quark masses at 2 GeV from \cite{pdg},
we get 
\begin{equation}
    \gzero \approx (-0.018 \pm 0.007)\,\btheta 
\end{equation}
and 
\begin{equation}
 \gone \approx (0.0012 \pm 0.0004)\,\btheta
\end{equation}
with $\sigma_{\pi N}(0) = 45\,{\rm MeV}$
and
$y  = 0.21 \pm 0.20$ from \cite{Borasoy_Meissner} as additional input.

Thus we find 
\begin{equation}
\frac{\gone}{\gzero} = -\frac{\epsilon^2}{2}\,\frac{M_\pi^2}{M_K^2\!-\! M_\pi^2}\,\frac{\sigma_{\pi N}(0)(1-y)}{\Mstr}
   \approx  
   -0.07 \pm  0.04
   \label{est_sigma}
\end{equation}
as the ratio of the isospin-breaking {\it versus} the isospin-conserving CP-violating $\pi NN$ coupling constants which
are induced by the $\theta$-term.
If we rather applied  the values $\sigma_{\pi N}(0) = 59(7)\,{\rm MeV}$ and $y\approx 0$ from
refs.~\cite{Oller1,MartinCamalich:2011py} (for an update of this work see ref.~\cite{ollernew}), we would get
\begin{equation}
 \gone \approx (0.0021 \pm 0.0004)\,\btheta
  \qquad \mbox{and} \qquad
   \frac{\gone}{\gzero} \approx
       -0.11 \pm 0.05
  \label{est_sigma_oller}
\end{equation}
 as values for $\gone$ and the ratio instead.
Note that the ratios listed in  (\ref{est_sigma}) and (\ref{est_sigma_oller})  are  compatible with the estimate (\ref{our_ratio}).

\setcounter{equation}{0}
\renewcommand{\theequation}{\thesection.\arabic{equation}} 

\section{Derivation via SU(3) chiral perturbation theory}
\label{app: su3}

In SU(3) ChPT
the $D$-type  and $F$-type CP-violating $\pi^0 NN$ coupling constants are (see  {\it e.g.} the U(3) ChPT calculation of
ref.~\cite{Ottnad_diplom})
\[
   g_{\pi^0 NN}^D = \frac{4 B \bar\theta \mstar}{F_\pi} b_D \qquad  \mbox{and}\qquad
    g_{\pi^0 NN}^F =   \frac{4 B   \bar\theta \mstar}{F_\pi} b_F\,,
\] 
respectively,
whereas 
\[ 
  g_{\eta NN}^D   =  \frac{-4 B \bar\theta \mstar}{F_\pi} \frac{b_D}{\sqrt{3}}\qquad  \mbox{and}\qquad  
   g_{\eta NN}^F  =  \frac{4  B   \bar\theta \mstar}{F_\pi}  \sqrt{3}b_F 
\] 
are the corresponding $\eta NN$ (actually $\eta_8 NN$) counter parts.
Note that $4B\, b_D$ and $4B\, b_F$ are the coefficients of the  anticommutator ($D$-type) and  commutator ($F$-type) term of the
quark mass matrix with the baryon matrix.
Therefore, the SU(3) counter parts of
eqs.~(\ref{gzero_form}) and  (\ref{gone_form})   are~\footnote{The proportionality of $\gone$ to $3b_F{-}b_D$ may come at first sight as a surprise.  Note, however, 
that the strange-quark content of the nucleon is  proportional to $b_0{+}b_D{-}b_F$ to leading order in the chiral expansion, such that $\gone$  for small or vanishing
$y$ is factually  proportional
to $2 b_0{+}b_D{+}b_F$ which in turn is proportional to $2c_1$. For more details see {\it e.g.} refs.\,\cite{MaiKubis,FrinkMeissner}.}
\begin{eqnarray}
\gzero &=&  \frac{4 B \btheta \mstar (b_D\!+\! b_F)} {F_\pi} 
 =   \btheta \frac{M_\pi^2}{F_\pi}(1\!-\!\epsilon^2) (b_D\!+\!b_F),\ \ \label{gzero_su3} \\
\gone &=& \frac{4 B\btheta \mstar}{F_\pi} \, \frac{3 b_F\!-\! b_D}{\sqrt{3}}\,\frac{\sqrt{3}}{4} \frac{m_d\!-\!m_u}{m_s \!- \!\hat m}
\nonumber \\ 
&=& 
 \bar\theta  \frac{M_\pi^2}{F_\pi}(1\!-\!\epsilon^2) \, (3b_F \!-\! b_D) \,\frac{ \epsilon\, M_\pi^2}{ 4(M_K^2-M_\pi^2) }\,, 
 \label{gone_su3}
\end{eqnarray}
where $(\sqrt{3}/{4}) (m_d-m_u)/(m_s - \hat m)$
is the $\pio$-$\eta$ (actually  $\pio$-$\eta_8$) mixing angle. 
Thus, in this case  we get  the ratio
\begin{equation}
   \frac{\gone}{\gzero} = \frac{\epsilon\,M_\pi^2}{4 (M_K^2- M_\pi^2)}\, \frac{ 3 b_F - b_D}{b_D+b_F} \,. 
\end{equation}
If  the values $b_F =-0.209\,{\rm GeV}^{-1}$ and
$b_D = 0.066\,{\rm GeV}^{-1}$ of ref.~\cite{BKM1995} are inserted,
we get the first row of table~\ref{tab: su3}.
Note, however, 
that there is a mismatch by  a factor  $1.5$ approximately between the SU(3) octet quantity
\[b_D+b_F = -\frac{m_\Xi-m_\Sigma}{4(M_K^2-M_\pi^2)}\approx (-0.143 \pm 0.004)\,{\rm GeV}^{-1} \]
used in \cite{Crewther,Pich,Ottnad}
and  the SU(2) low-energy coefficient (LEC)
\begin{equation} 
 c_5 = \frac{\Mstr}{4 \mpi^2 \epsilon}
 \approx 
 (-0.097\pm 0.034)\,{\rm GeV}^{-1} \,,
  \label{eq: c5_value}
\end{equation}
although according to SU(3) ChPT 
both quantities should agree to leading order, see eq.~(27) of ref.~\cite{MaiKubis}\footnote{In fact, the latter equation
which is based on eq.~(5.7) of ref.~\cite{FrinkMeissner} predicts 
that the \NLO\ correction to
$c_5$ is much larger than $c_5$ (or $b_D+b_F$) itself, namely
$\Delta c_5 = 0.49\, {\rm GeV}^{-1}$. This quantity is of similar size
as $\Delta c_1 = + 0.2\,{\rm GeV}^{-1}$. \label{foot:c5}}.

Moreover, an alternative procedure  to parametrize  the above sum is 
\begin{eqnarray}
   b_D+b_F &=& \frac{\Mstr}{4 (M_{K^+}^2-  (M_{\pi^+}^2- M_{\pi^0}^2) -M_{K^0}^2)} \nonumber\\
    &\approx& 
    (-0.126 \pm 0.024)\,{\rm GeV}^{-1}\,,
    \label{eq:bdbfiso}
\end{eqnarray}
where the electromagnetic mass shifts are removed (via the Dashen theorem\,\cite{Dashen} in the denominator) and where the prediction falls 
in-between the original one and the $c_5$ value.
Using an analogous parametrization for $b_F$, we get
\begin{equation}
   b_F = \frac{ m_{\Sigma^-} - m_{\Sigma^+}}{8 (M_{K^+}^2 - (M_{\pi^+}^2 - M_{\pi^0}^2) - M_{K^0}^2)}
   \approx -0.196 \,{\rm GeV}^{-1}
   \label{eq:bfnew}
\end{equation}
and 
$b_D = +(0.069\pm 0.024)\,{\rm GeV}^{-1}$  from (\ref{eq:bdbfiso}) 
instead of the above listed values 
from \cite{BKM1995}, such that  the values in the second row of table~\ref{tab: su3} are generated instead.
Note that the result for  $3 b_F - b_D$ is approximately the same in both parametrizations, 
namely $-0.69\,{\rm GeV}^{-1}$ in the original one \cite{BKM1995}  and  
 $-0.66\,{\rm GeV}^{-1}$ in the modified one.

Finally, replacing $b_D+b_F$ of \cite{BKM1995}
by $c_5$ of eq.~(\ref{eq: c5_value}), we get the values in the third row of   table~\ref{tab: su3}.

Note that only the last SU(3) value of the ratio $\gone/\gzero$ is in the range of our estimate (\ref{our_ratio}), but
all three are compatible with the estimate of (\ref{est_sigma}).  The quoted numbers of table~\ref{tab: su3}, however, do not contain a systematical
error connected with an SU(3) ChPT calculation. For standard quantities such an uncertainty  is certainly of the order of 50\,\%. For the
quantity $c_5$ this uncertainty should be rather 100\,\%---200\,\%, see {\it e.g.} footnote\,\ref{foot:c5}.  Taking
these SU(3) errors into account, the estimates of table~\ref{tab: su3} are compatible with the range quoted in (\ref{our_ratio}).

\setcounter{equation}{0}
\renewcommand{\theequation}{\thesection.\arabic{equation}} 

\section{The contribution of the odd-parity nucleon resonance to $\gone$}
\label{app: odd parity}

\begin{figure} 
 	\centering
  \includegraphics[scale=0.45]{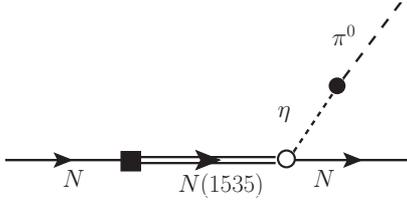}
  \caption[odd parity contribution]
  {Effective CP-violating and isospin-violating $\pi_3 NN$ vertex estimated as
   CP-violating transition  (black square) from the even-parity nucleon to the
   odd-parity  $S_{11}(1535)$ nucleon-resonance (double line) which in turn decays into $\eta N$ (open circle) with subsequent isospin-breaking 
   by
  $\eta-\pio$ mixing (black circle).  Note that the second topology of the diagram where the pion emission comes first is included in the
  calculation.
  \label{fig: odd parity}}
\end{figure}

According to ref.\,\cite{pdg} the mass, width and $N\eta$ branching ratio  of the $S_{11}(1535)$ odd-parity
nucleon-resonance are 
$m_{N_{1535}}=  (1535 \pm 10)\mev$,  $\Gamma_{N_{1535}}= (150\pm 25)\mev$ and ${\cal B}_{N_{1535}\to N\eta} = (42\pm 10)\,\%$. 
Finally the CM-momentum is $p^{\star} = 186\mev$.
The partial decay width $\Gamma_{N_{1535} \to N\eta}$
is then  approximately $63\mev$, such that one finds for the effective
coupling constant for the decay $N^*\to N\eta$
 \begin{equation}
  | g^{\star} | =  \sqrt {\frac{8\pi \Gamma_{N_{1535} \to N\eta }} {p^{\star}} }\approx 2.9\,,
  \end{equation}
  where we assumed an energy-independent decay vertex.
 By inserting
 \begin{equation}
  \half \langle i \chi_-\rangle = -\mpi^2(1{-}\epsilon^2)\btheta (1 - \half \pi^2/F_\pi^2)+ \epsilon 2 \mpi^2 \pi_3/F_\pi + \dots
\end{equation}
into the effective   interaction Lagrangian 
\begin{equation}
  {\cal L}_{N_{1535} N}  =      \tilde h {N^\dagger_{1535}} \half \langle i\chi_-\rangle N + \mbox{h.c.}
\end{equation}
   we get
  \begin{equation}
  {\cal L}_{N_{1535} N} 
      =       \tilde h N^\dagger_{1535}\! \left (\!- M_\pi^2(1{-}\epsilon^2)\bar\theta +\epsilon\frac{2 M_\pi^2}{F_\pi}  +  \dots\!\right)\!N
         + \mbox{h.c.} 
                 \label{Lag_NNstar}
 \end{equation}
The first term provides the CP-odd transition of a nucleon into the $N^*$.
As illustrated in fig.\,\ref{fig: odd parity},  we may model  the second vertex by the decay of the resonance into an $\eta$ and a
nucleon, followed by  $\eta$--$\pio$ mixing; using the leading order
ChPT expression for the mixing amplitude
  \begin{equation}
   \epsilon_{\pi^0\eta}
                                     \approx {\sqrt{\third}}
                                     \,\frac{B(m_d-m_u)}{M_{\eta}^2 - M_\pi^2}
                                     \approx {1.37} \, \%\,, \label{eps_eta}
     \end{equation}
  we can express $\tilde h$ 
 by  $g^\ast$ and $\epsilon_{\pi^0\eta}$ as
 \begin{equation}
  \tilde h = \epsilon_{\pi^0\eta}\frac{F_\pi g^\star}{2\epsilon M_\pi^2} \,.
 \end{equation}
Thus the interaction Lagragian (\ref{Lag_NNstar}) can be rewritten 
 as
 \begin{equation}
  {\cal L}_{N^\star N\chi_-} 
                                                = g^\star \epsilon_{\pi^0\eta} \left( \frac{F_\pi (1{-}\epsilon^2)\btheta}{-2\epsilon} + \pi_3  \right) N^\ast_{1535} N   + \mbox{h.c.}\, 
 \end{equation}
In summary, we get  the following estimate for the odd-parity contribution to the CP-violating isospin-breaking $\pi NN$ coupling constant
 \begin{equation}
  \delta \gone =  |g^\star|^2 (\epsilon_{\pi^0\tilde\eta})^2\,\frac{\btheta F_\pi(1{-}\epsilon^2)/(-\epsilon)}{m_{N_{1535}}-m_N} 
                        \approx  (0.6\pm 0.3)\cdot 10^{-3}\,\btheta\,,
 \end{equation}
 which is only one third of the NDA estimate 
 \[ |\epsilon| \frac{\mpi^4}{m_N^3 \fpi}\,\btheta \sim 1.7 \cdot 10^{-3}\,\btheta\,. \]

\end{appendix}



\end{document}